\documentclass[review]{elsarticle}
\usepackage{graphicx}
\usepackage{lineno,hyperref}
\usepackage{dcolumn}
\usepackage{bm}
\usepackage{epsfig}
\usepackage{caption}
\usepackage[english]{babel}
\usepackage[english]{babel}
\usepackage{amsmath}
\usepackage{tikz}
\usepackage{multicol}

\modulolinenumbers[5]

\begin{document}

\begin{frontmatter}

\title{Real-time Diffusion Monte Carlo method}

\author{Ilkka Ruokosenm\"aki, and Tapio T.~Rantala}
\address{Department of Physics, Tampere University of Technology, Finland}
\ead{ilkka.ruokosenmaki@tut.fi, Tapio.T.Rantala@iki.fi}

\begin{abstract}

Direct sampling of multi-dimensional systems with quantum Monte Carlo methods allows exact account of many-body effects or particle correlations.  The most straightforward approach to solve the Schr\"odinger equation, Diffusion Monte Carlo, has been used in several benchmark cases for other methods to pursue.  Its robustness is based on direct sampling of a positive probability density for diffusion in imaginary time.  It has been argued that the corresponding real time diffusion can not be realised, because the corresponding oscillating complex valued distribution can not be used to drive diffusion.   Here, we demonstrate that this can be done with a couple of tricks turning the distribution piecewise positive and normalisable.  This study is a proof of concept demonstration using the well-known and transparent case: one-dimensional harmonic oscillator.  Furthermore, we show that our novel method can be used to find not only the ground state but also excited states and even the time evolution of a given wave function.  Considering fermionic systems, this method may turn out to be feasible for finding the wave function nodes.


\end{abstract}

\begin{keyword}
Path integral, quantum dynamics, first-principles, Monte Carlo, Real-Time
\end{keyword}

\end{frontmatter}

\pagebreak


\section{Introduction}

Quantum Monte Carlo (QMC) methods form a collection of robust approaches to study quantum many-particle systems.\cite{Thijssen}  With QMC the central benefit is that one can deal with multi-dimensional systems, where standard grid based methods become computationally too heavy. With Path Integral and Green's function approaches the  many-body effects or correlations can be taken into account without introducing approximations and evaluated within numerical accuracy, which is limited by the computational resources, only.  Furthermore, if starting from the first-principles, also the systematic errors are avoidable.  Thus, for the field of electronic structure calculations, with QMC one can benchmark the energetics, structure and dynamics of atoms and molecules with desired accuracy.  It is even straightforward in cases where the wave function is everywhere positive or can be considered as piecewise positive between given nodes.

Diffusion Monte Carlo (DMC) or Green's function Monte Carlo is a typical representative of QMC.  In several cases it has been demonstrated to be a simple but accurate approach to find the ground state \cite{Thijssen,DMCbook}.  In particular, both bosonic \cite{KalosCeperley} and fermionic \cite{KalosOld,DMC-96} systems have been successfully considered.  A recent example is benchmarking the hydrogen molecule and its simple reaction conformations with increasing accuracy \cite{Anderson}.  

With DMC the Schr\"odinger equation in imaginary time turns to a diffusion equation, whose "imaginary time evolution" or iteration converges to the ground state.  Transformation of the Schr\"odinger equation to the corresponding integral equation shows how  diffusion can be simulated with random walkers guided by the interactions of quantum particles.  The walker distribution, which is everywhere positive converges to the ground state wave function.  This is the simple idea of DMC simulation, where it is essential that the product of the wave function and diffusion probability is everywhere positive.  The latter one is the kernel of the integral equation \cite{Anderson, feynman,paper1,paper2}.

Due to the everywhere positive "diffusion distribution" interpretation as the wave function, simulation of excited states and indistinguishable fermions becomes problematic with DMC \cite{KalosOld, KalosNew}.  Nodes of the wave function should be known, {\it e.g.}~by symmetry, or approximated with good enough accuracy to make it piecewise positive.  Though there are practical approximate ways around the problem, mostly with approximate nodes, this remains as an impediment with DMC.

Based on the probability interpretation of the kernel and wave function, and diffusion nature of the random walk, it has been argued that the simple and useful principles of DMC, above, can not be used to solve the Schr\"odinger equation with real time path integrals \cite{makri, Kieu}.  In this study we show that this is not true and we present a practical solution to this problem, which is related to the sc.~"numerical sign problem" of real-time path integrals.  Furthermore, we demonstrate that our new real-time path integral approach is capable of finding, not only the ground state, but also excited states, and also, it can be used to simulate proper real time quantum dynamics -- not to be mixed with diffusion.

This study is a proof-of-concept demonstration of a novel "real-time DMC".  Therefore, we have chosen a transparent test case, the well-known one-dimensional harmonic oscillator (ODHO), where the method and its performance are clearly seen.  We also benefit from the exact propagator of the harmonic oscillator while the testing the real-time diffusion.

\section{Diffusion Monte Carlo and its real time counterpart}

\subsection{Positive probability density}

The imaginary time \( (\tau = it) \) integral equation of the conventional Diffusion Monte Carlo (DMC or $\tau$DMC) for the many-body wave function \(\psi\) is
\begin{equation} \label{imaginary}
\psi(x_b,\tau_b) = \int_a G(x_b,\tau_b;x_a,\tau_a)\psi(x_a,\tau_a){\rm d} x_a,                
\end{equation}
where the kernel \( G \) is the Green's function of the system, the position space re\-presentation of the imaginary time evolution operator.  We present the formalism in one-dimensional space of \(x\), here, and in what follows, but extensions to more dimensions is trivial.  For a time step \( \tau = \tau_b - \tau_a \), and
\begin{equation} \label{DMCkernel}
\begin{split}
G \approx \ & G_{\rm{diff}}G_{\rm{B}},                              \\                                                              
& G_{\rm{diff}} = C \exp\left[ -(x_b-x_a)^2/2 \tau \right] , \\  
& G_{\rm{B}} = \exp\left[ -\left(\frac{1}{2}(V(x_b)+V(x_a)) - E_T\right) \tau \right] , 
\end{split}
\end{equation}
where $C = (2 \pi \tau)^{-3/2}$ and $E_T$ is the trial energy, iterated to the ground state total energy at self-consistency, \( \psi(x_b) = \psi(x_a) \).  The Green's function and the stationary solution of Eq.~(\ref{imaginary}) becomes exact as \( \tau \rightarrow  0 \).

Now, the kernel \( G \) is everywhere real and positive, and therefore, it can be considered as a normalizable probability density in Monte Carlo evaluation of the ground state wave function \( \psi(x) \) as the stationary walker density \cite{DMCbook}.  The power of $\tau$DMC arises from the independence of Monte Carlo walkers in "diffusion", and also, the locality of \( G_{\rm{diff}} \), which increases the accuracy of  \( G_{\rm{B}} \).


For the real time propagation of the many-body wave function \( \psi(x,t) \) there is a similar integral equation \cite{feynman}
\begin{equation}   \label{realtime}
	\psi (x_b, t_b) = \int_a  K(x_b,t_b;x_a,t_a) \psi (x_a, t_a) {\rm d} x_a,                  
\end{equation}
where the kernel \( K \) is the path integral over the time step \( t =  t_b - t_a \), (\(t_a < t_b\)),
\begin{equation}    \label{kernel}
	K(x_b,t_b;x_a,t_a) = \int_{x_a}^{x_b} \exp( {\rm i} S[x_b,x_a] ) \mathcal{D}x(t).       
\end{equation}
Here \(S[x_b,x_a] = S[x](x_b,x_a) = \int_{t_a}^{t_b} L_x dt  \) is the action of the path \(x(t)\) from \(a=(x_a,t_a)\) to \(b=(x_b,t_b)\) and \( L_x \) is the corresponding Lagrangian \cite{feynman}.   Now, finding the Monte Carlo solutions for \( \psi \) from Eqs. (\ref{imaginary}) and (\ref{realtime}) greatly differ.


The $\tau$DMC diffusion like procedure can not be used to solve Eq.~(\ref{realtime}) for \( \psi \), because the kernel \( K \), as a path integral, is a complex valued functional of interfering paths coupling all of the walkers.  Thus, \( K \) can not be interpreted as a probability \cite{makri, Kieu}, and furthermore, it is delocalised with complex exponential tails oscillating in whole space, the more the shorter the time step $t$.

Here, we present a novel idea solving this problem and formulate a "real-time diffusion Monte Carlo" ($t$DMC or RTDMC) procedure, which retains the advantage of "diffusion of independent walkers".  Furthermore, the $t$DMC enables evaluation of excited states and even real time quantum dynamics, out of reach with the $\tau$DMC.  We have these advanced features in our direct real-time path integral (RTPI) approach \cite{paper1,paper2}, already, but there, all of the paths coupling the walkers $ \{ x_{ai} \}_{i=1}^{N_a} $ and $ \{ x_{bj} \}_{j=1}^{N_b} $ need to be considered.  With increasing number of walkers it leads to quadratic growth (\( \propto N^2 \), assuming  $N_a = N_b = N$) of computational efforts with RTPI.  With $t$DMC, however, the growth of efforts is close to linear (\( \propto N \)), only.


First, we separate the integrand in Eq.~(\ref{realtime}) to terms, which can be considered as "positive probabilities", and second, we accomplish normalization by restricting the space of integration.  
We separate similarly both the kernel \( K \propto \exp({\rm i}\phi) \) \cite{feynman} and the wave function \( \psi(a) \) at the right hand side of (\ref{realtime}) to four parts as
\begin{equation} \label{Kern}			                               					
\begin{split}
K (b, a) & = C \exp({\rm i}\phi) = C \left[\cos(\phi) + {\rm i} \sin(\phi) \right ] = C \left[\cos(\phi) + {\rm i} \cos(\frac{\pi}{2} - \phi) \right ]  \\	
& = C \left[\cos^2(\frac{\phi}{2}) - \sin^2(\frac{\phi}{2}) + {\rm i} \left(\cos^2 \left(\frac{\frac{\pi}{2} - \phi}{2} \right) - \sin^2 \left(\frac{\frac{\pi}{2} - \phi}{2} \right)\right) \right ]  \\		
& = K_{+} (b, a) - K_{-} (b, a) + {\rm i} K_{+\rm i} (b, a) - {\rm i} K_{-\rm i} (b, a) 
\end{split}     
\end{equation}
and
\begin{equation} \label{psi}
\psi (a) = \psi_{+} (a) - \psi_{-} (a) + \rm i \psi_{+\rm i} (a) - \rm i \psi_{-\rm i} (a).               
\end{equation}
This splits the integrand into 16 terms.  Here \(C\) and $\phi$ are some functions of \(a\) and \(b\), that can be chosen so that \(C\) is real and positive. Rearrangement of these terms allows splitting the left hand side of (\ref{realtime}) with the same principle as
\begin{equation} \label{four}			                               					
\begin{split}
& \psi_+(b) = \int_a K_+ \psi_+ dx_a + \int_a  K_- \psi_- dx_a+\int_a K_{+i}\psi_{-\rm i}dx_a+\int_a K_{-i}\psi_{+\rm i}dx_a  \\
&  \psi_-(b) = \int_a K_+ \psi_- dx_a + \int_a K_- \psi_+dx_a +\int_a K_{+i}\psi_{+\rm i}dx_a+\int_a K_{-i} \psi_{-\rm i}dx_a  \\
&  \psi_{+ \rm i}(b) = \int_a K_+  \psi_{+\rm i}dx_a + \int_a  K_-  \psi_{- \rm i}dx_a+\int_a K_{+i} \psi_{+}dx_a+\int_a K_{-i} \psi_{-}dx_a  \\
&  \psi_{- \rm i}(b) = \int_a K_+  \psi_{-\rm i}dx_a + \int_a K_-  \psi_{+ \rm i} dx_a+\int_a K_{+i} \psi_{-} dx_a+\int_a K_{-i} \psi_{+}dx_a  ,
\end{split}     
\end{equation}
each of which is everywhere real and positive. Here, all of the \( K_{\rm sub} \) and \( \psi_{\rm sub} \) on the right-hand side stand for \( K_{\rm sub}(b, a) \) and \( \psi_{\rm sub}(a) \), respectively, where \(a=(x_a,t_a)\), \(b=(x_b,t_b)\) and sub = \{ $ +, -, +i, -i�$ \}.
Thus, the complete wave function at the end of the time step \( t = t_b - t_a \) can be written as
\begin{equation} \label{psbi}
\psi (b) = \psi_{+} (b) - \psi_{-} (b) + \rm i \psi_{+\rm i} (b) - \rm i \psi_{-\rm i} (b).               
\end{equation}

Thus, our approach is reminiscent of an old $\tau$DMC method of Arnow {\it et.al.} \cite {KalosOld}, where positive and negative walkers were used for the respective parts of the wave function.  The main differences are the following.  Here, we have four types of walkers and each walker generates all other types of walkers.  Therefore, all parts of Eqs.~(\ref{four}) are correlated and unlike in $\tau$DMC  \cite {KalosOld} they do not separately converge to the ground state, but instead, we are able to simulate time evolution of a complex time-dependent wave function, as discussed below.



In Eqs.~(\ref{four}), we have a fully delocalised piecewise everywhere positive probability density to sample, which first needs to be normalised.  In case of a wave function localized in a finite domain we know that the contributions to \( \psi (b) \) in Eq.~(\ref{psbi}) cancel outside the domain and close to the domain boundaries inside.  Then, we can normalise the partial probabilities of Eq.~(\ref{Kern}) in a so chosen domain and run diffusion localised in the domain.  Next, let us discuss the kernel and related approximations.

\subsection{Kernel}

The kernel in closed form is known for a few special systems, only \cite{feynman, HOKernel}.  The harmonic oscillator ($ V(x) = {1 \over 2} m \omega^2 $) is one of those with the kernel
\begin{equation} \label{HOkernel}                                                                                    
\begin{split}
K(x_b,t_b;x_a,t_a) = & \exp(-i\theta) \left(\frac{m \omega}{2 \pi \hbar |\sin(\omega t)|} \right)^{1/2} \times \\
& \exp \{\frac{i m \omega}{2 \hbar \sin(\omega t)}[(x_b^2 + x_a^2)\cos(\omega t) - 2 x_b x_a] \}, 
\end{split}  
\end{equation}
where $t = t_b - t_a$ and $\theta = \frac{\pi}{4}(1+2 \, \rm{trunc}(\omega t/\pi))$. Here, "trunc$(x)$" denotes the truncation function, the largest integer less than or equal to $x$.

In general, for a given potential $�V(x)�$ we need to approximate kernels and the most usual approximation is sc.~"short time approximation" or Trotter kernel \cite{schulman, makri}
\begin{eqnarray}  \label{trottpropag}
 K(x_b,t_b;x_a,t_a) \approx \left[\frac{1}{2 \pi {\rm i}  t}\right]^{N/2} \exp \left[\frac{\rm i}{2  t} ( x_b - x_a )^2 - \frac{{\rm i}  t}{2}(V(x_a) + V(x_b)) \right] ,                                                                                                 
\end{eqnarray}
which becomes exact as \( t \rightarrow 0 \), cf.~Eq.~(\ref{DMCkernel}).


Both of the kernels (\ref{HOkernel}) or (\ref{trottpropag}) can be written in the piecewise positive form by using the  recipe given in  Eq. (\ref{Kern}).  For the Trotter kernel we define notations: average Lagrangian  \(\bar{L} = \left[\frac{1}{2 t} ( x_b - x_a )^2\right] - \left[ \frac{t}{2}(V(x_a) + V(x_b)) \right]\),  \(C =  \left[\frac{1}{2 \pi  t}\right]^{1/2} \) and \(D =  \frac{C\sqrt{2}}{2}\).  Then, we write
\begin{equation}  \label{a}
\begin{split}
 & K(b,a)  = C(-{\rm i})^{1/2} \exp({\rm i}(\bar{L})) = C \exp({\rm i}(\bar{L}-\frac{\pi}{4}))  \\
 & = \frac{C\sqrt{2}}{2} \left[ \cos(\bar{L}-\frac{\pi}{4}) + {\rm i} \sin(\bar{L}-\frac{\pi}{4}) \right]  
  = D\left[ \cos(\bar{L}-\frac{\pi}{4}) + {\rm i} \cos (\frac{3\pi}{4}-\bar{L}) \right]  \\
 & = D\left[\cos^2(\frac{\bar{L}-\frac{\pi}{4}}{2}) - \sin^2(\frac{\bar{L}-\frac{\pi}{4}}{2}) + {\rm i}(\cos^2(\frac{\frac{3\pi}{4} -\bar{L}}{2}) - \sin^2(\frac{\frac{3\pi}{4}-\bar{L}}{2}) )\right]  \\
& \equiv D \left[ K_{+}(b,a) - K_{-}(b,a) + {\rm i} K_{+ \rm i}(b,a)- {\rm i} K_{-\rm i}(b,a)\right] .     
\end{split}                                                                             
\end{equation}

In case of the harmonic oscillator it should be noted, that while the accuracy of short time approximation increases with decreasing time step, the exact kernel allows any length of time step.  However, both of these kernels diverge for \( t  = 0\) and the exact one also periodically for \( t_n = n \pi / \omega \) .

\subsection{Real-time diffusion}


While the imaginary time diffusion is a very local phenomenon, the more the shorter the time step \( \tau \), whereas, the real-time diffusion is fully delocalized in form of oscillatory  \( \sin^2 \) and \( \cos^2 \) functions, the wave length depending on the average Lagrangian in the time step \( t \).  Thus, it is sufficient to consider and normalize these distributions in the chosen domain, only, and correctly with respect to each other.  Diffusion out of the box can be ignored, because it is known that the different contributions in Eqs.~(\ref{psbi}) cancel at long distances.




The four parts of the initial wave function \( \psi(a) \) in Eq.~(\ref{psi}) are presented with corresponding four sets of walkers, whose total number is $ N_a�$. Neither real contributions \( \psi_+(a) \) and \( \psi_-(a) \) nor the imaginary contributions \( \psi_{+i}(a) \) and \( \psi_{-i}(a) \) should pairwise overlap as the complex wave function should be single valued.  This is not absolutely necessary to carry on calculations, as we show later.  Now, the real-time diffusion of these walkers according to the Eq.~(\ref{four}) results in four strongly delocalised and pairwise overlapping contributions, real \( \psi_+(b) \) and \( \psi_-(b) \), and imaginary  \( \psi_{+i}(b) \) and \( \psi_{-i}(b) \).  To render the wave function \( \psi(b) \) in Eq.~(\ref{psbi}) single valued, the pairwise overlap should be removed.  This is carried out by cancellation or pairwise annihilation of nearby walkers until the nodal surfaces between the positive and negative amplitudes appear.

There is a large cancellation of walkers also in the box, {\it e.g.,} the wave function must vanish close to the domain boundaries, and similar cancellation turns out to dominate everywhere in the domain.  In fact, it is only a small fraction of walkers, which eventually remain presenting the wave function.  Due to the massive cancellation of diffusing walkers all initial walkers need to be massively duplicated in each time step to maintain the total number of walkers.

\begin{figure}
\centering
\begin{tikzpicture}

\node at (-9,0) {\includegraphics[trim=3cm 5cm 0.1cm 8cm, clip=true,scale=0.43]{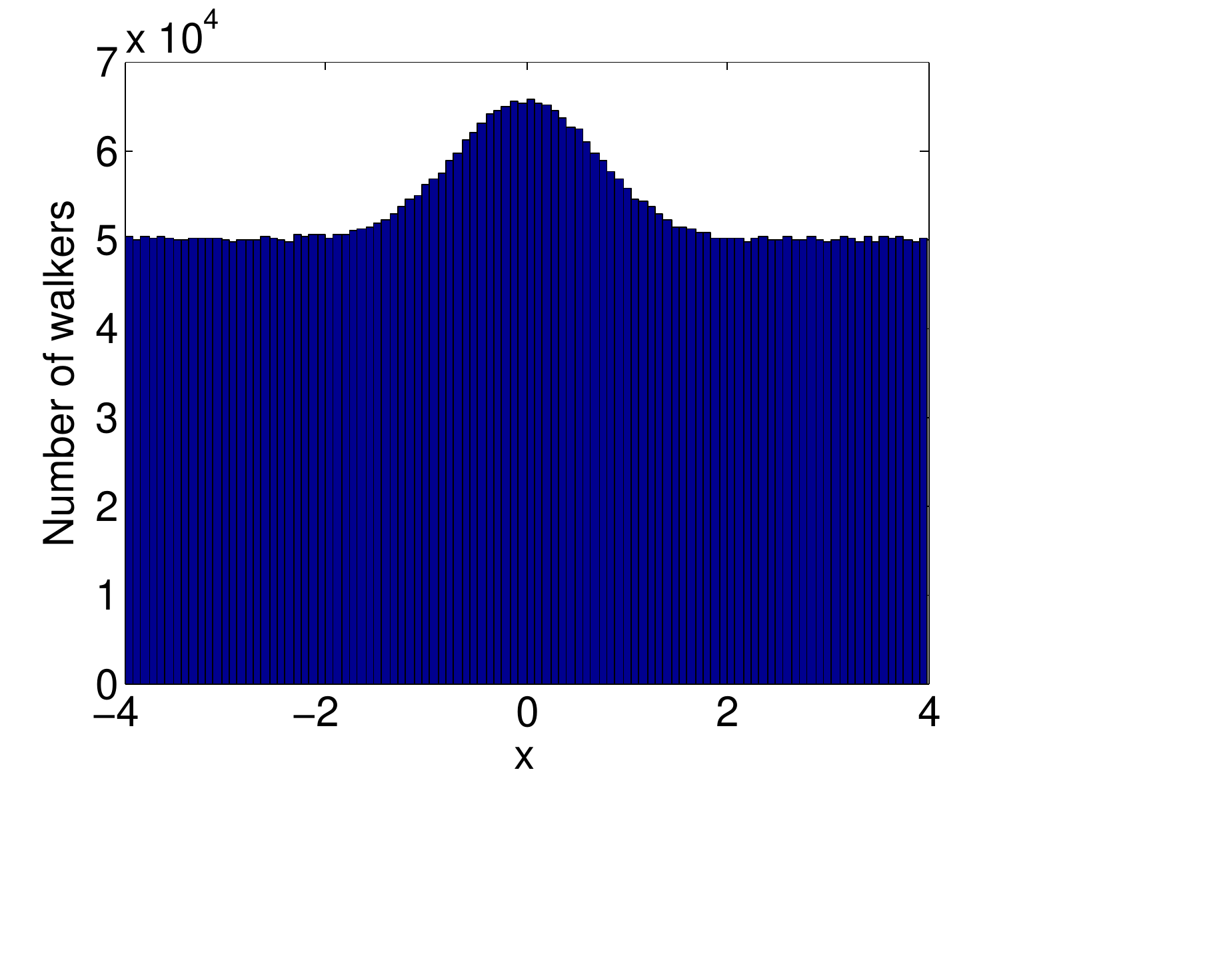}};
\node at (-2,0) {\includegraphics[trim=3cm 5cm 0.1cm 8cm, clip=true,scale=0.43]{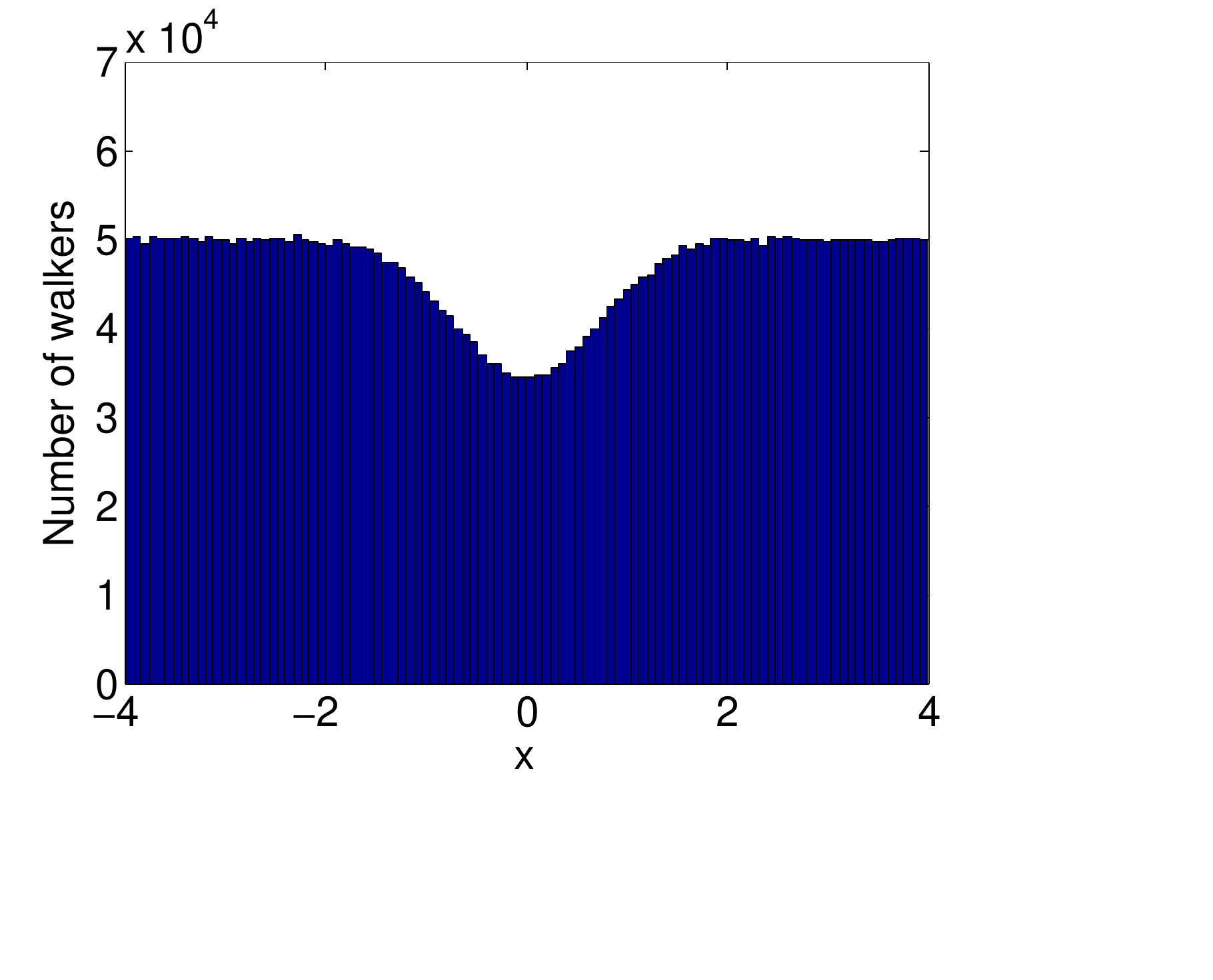}};

\node at (-11.5,2.1) {a)};
\node at (-4,2.4) {b)};

\end{tikzpicture}
\vskip -1.5cm
\caption{Distribution of a) positive and b) negative walkers (\( \psi_+(b) \) and \( \psi_-(b) \)) after one time step $t = 0.1$ from gaussian real wave function \( \psi_+(a) \) and $N(x_a) \approx 10^7$ walkers.  Histogram bin width is 0.08.}
\label{XP&XN}
\end{figure}

A one-timestep real time diffusion is demonstrated in Fig.~\ref{XP&XN}.  The initial state is ODHO ground state gaussian real wave function, {\it i.e.}, \( \psi(a) = \psi_+(a) \).  The real components \( \psi_+(b) \) and \( \psi_-(b) \) after propagation with the exact kernel (\ref{HOkernel}) over a short time step \( t \) are shown.  We see that most of the walkers will cancel out, leaving behind the initial real gaussian shape, but slightly scaled down.  Similarly, the \( \psi_{+i}(b) \) and \( \psi_{-i}(b) \) after cancellation result in a small negative gaussian shape for the imaginary part, as expected, not shown in Fig.~\ref{XP&XN}.  This corresponds to rotation of the wave function from the real axis downwards with a small angle, which is interpreted as multiplication with the phase factor $�{\rm e}^{-iEt/\hbar}�$.


Here we use a simple one-dimensional cancellation algorithm.  We define a {\sl walker touch parameter} $\delta$, and when positive and negative walkers appear closer than $\delta$, they annihilate each other. Finding an efficient cancellation algorithm turns out to be a key factor in the present method with large number of walkers and oscillatory nature of tDMC propagators it may become a key issue in multidimensional spaces.  Continuation without walker annihilation leads to waste of efforts, as can be predicted from Fig.~\ref{XP&XN}, and finally, losing the remaining meaningful wave function into noise. This is one manifestation of the "sign problem", which still is an area of ongoing research \cite{Anderson, KalosNew, Alexandru}.

\section{Coherent propagation}

First, we consider straightforward simulation of quantum dynamics by using the above developed tDMC.  We call this {\sl coherent propagation}, because the phase factor of the wave function is properly treated.  Next, we consider {\sl incoherent propagation} and demonstrate its use for finding the stationary eigenstates of the system instead of running full quantum dynamics.

\subsection{Quantum dynamics from real time diffusion}

Because this study is a "proof of the concept tDMC", we continue with the simple, well-known and transparent ODHO as the test test bench.  Furthermore, for ODHO we have the exact propagator available, and thus, the issues related with the real time diffusion and approximate propagators can be investigated separately.

Hence, we run dynamics of a particle in potential \( V(x) = {1 \over 2} m \omega^2 x^2 \) with \( \omega = 2 \).  This may be related to an electron in a "harmonic quantum dot" or in an atom.  Thus, it is practical to use related atomic units, where $m = \hbar = a_0 = 1$, where $ a_0 $ is the Bohr radius and the unit of time is $ (m a_0^2) / \hbar \approx 24$ as.  Now,  $ \omega = 2$ corresponds to relatively strong confinement.

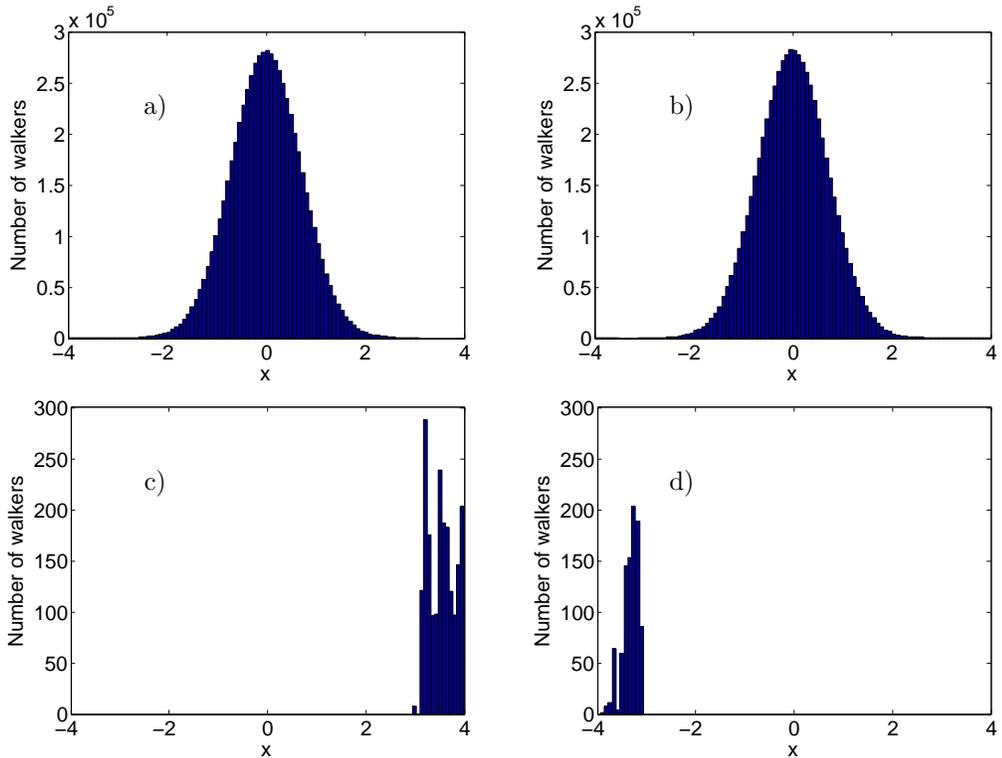
\begin{figure}
\vskip-3.5cm
\centering
\begin{tikzpicture}

\node at (-9,0) {\includegraphics[trim=3cm 5cm 0.1cm 8cm, clip=true,scale=0.43]{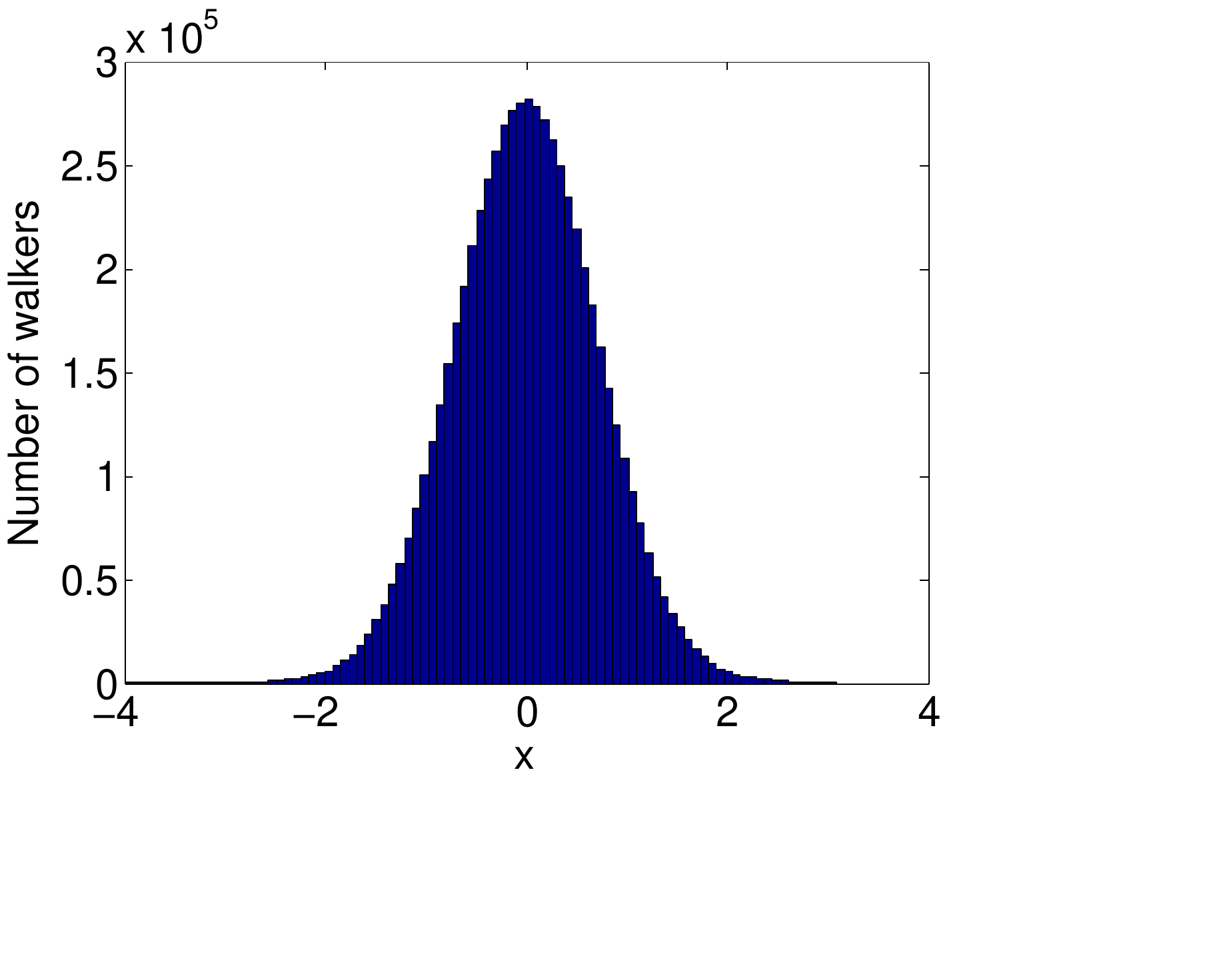}};
\node at (-2,0) {\includegraphics[trim=3cm 5cm 0.1cm 8cm, clip=true,scale=0.43]{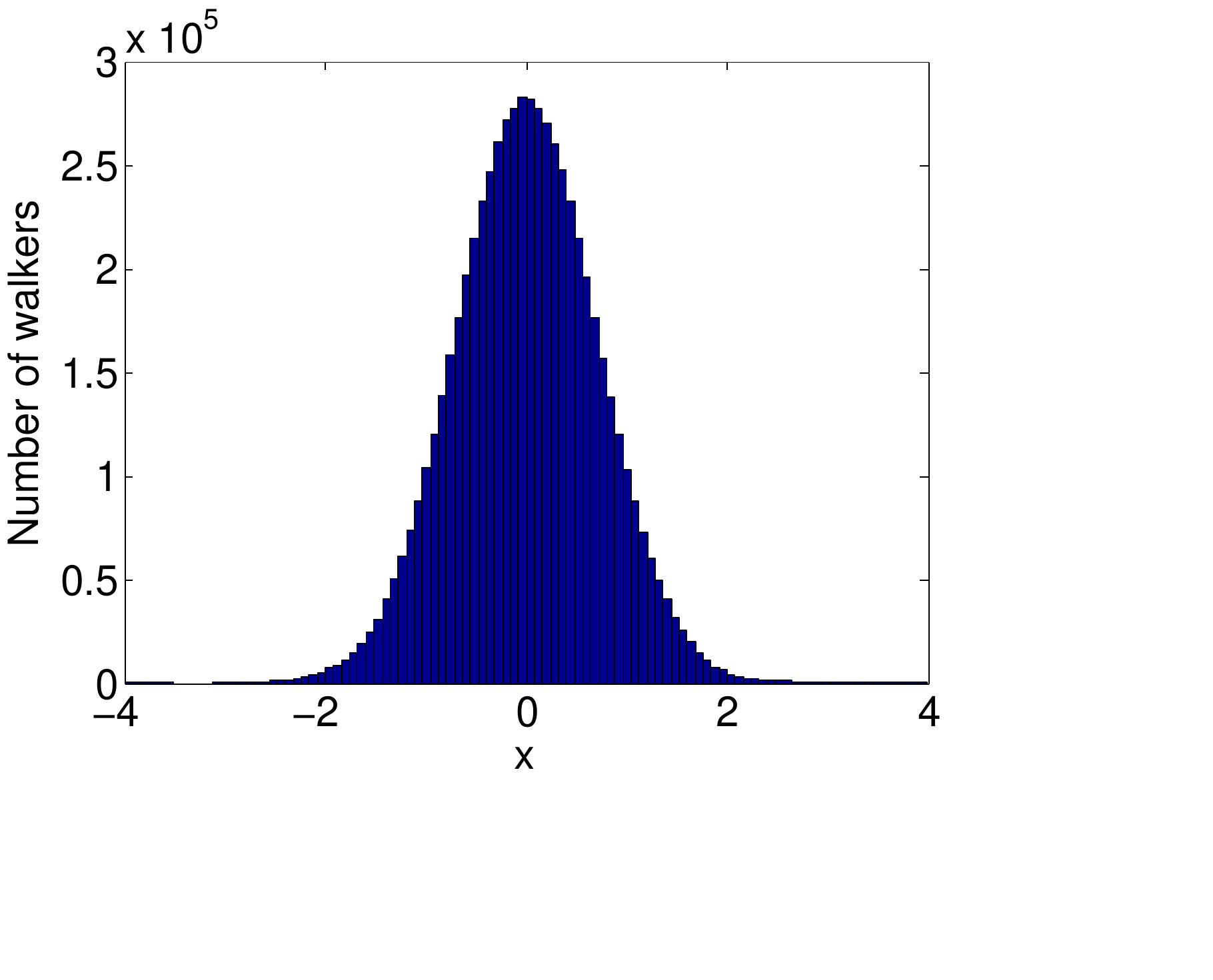}};
\node at (-9,-5) {\includegraphics[trim=3cm 5cm 0.1cm 8cm, clip=true,scale=0.43]{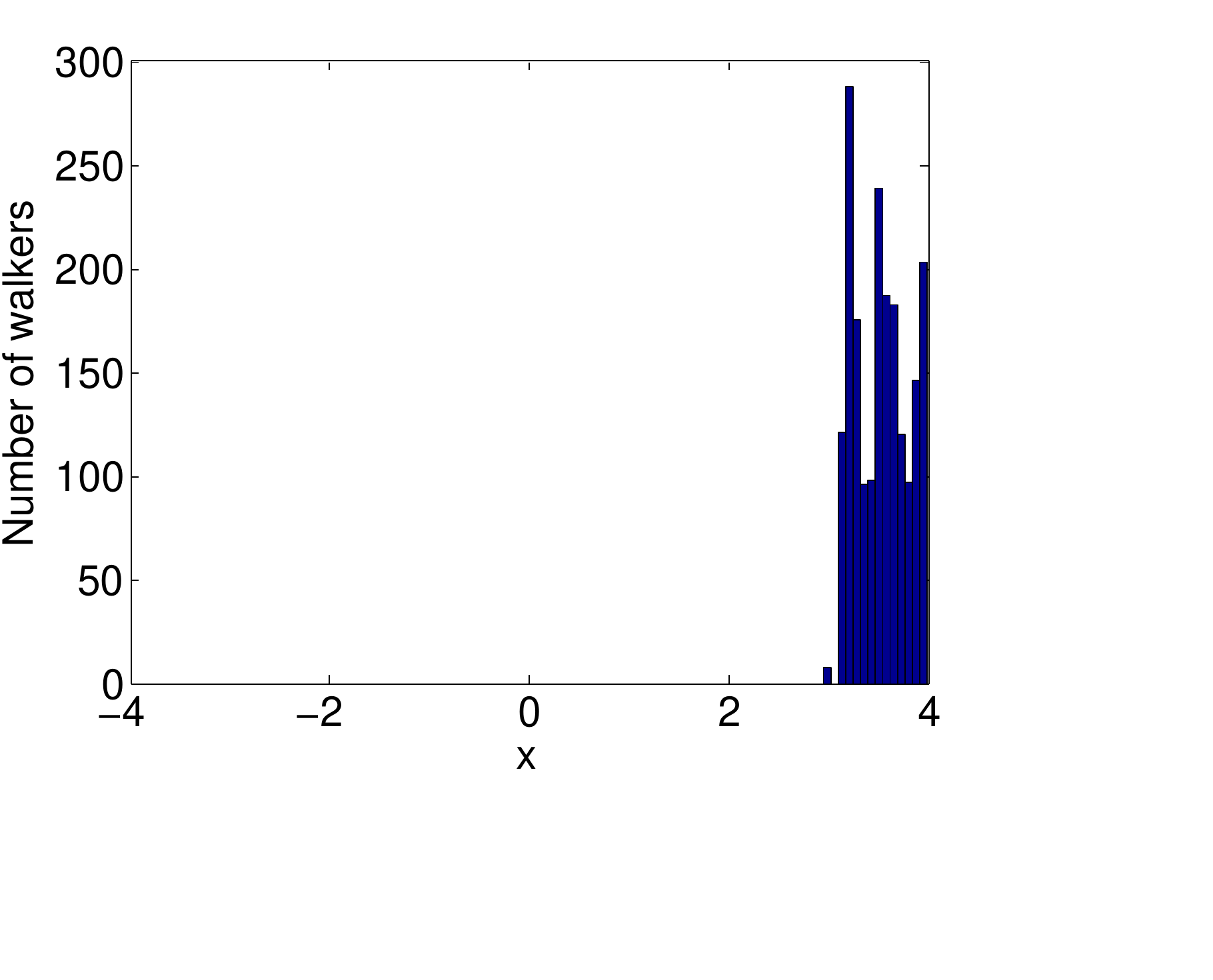}};
\node at (-2,-5) {\includegraphics[trim=3cm 5cm 0.1cm 8cm, clip=true,scale=0.43]{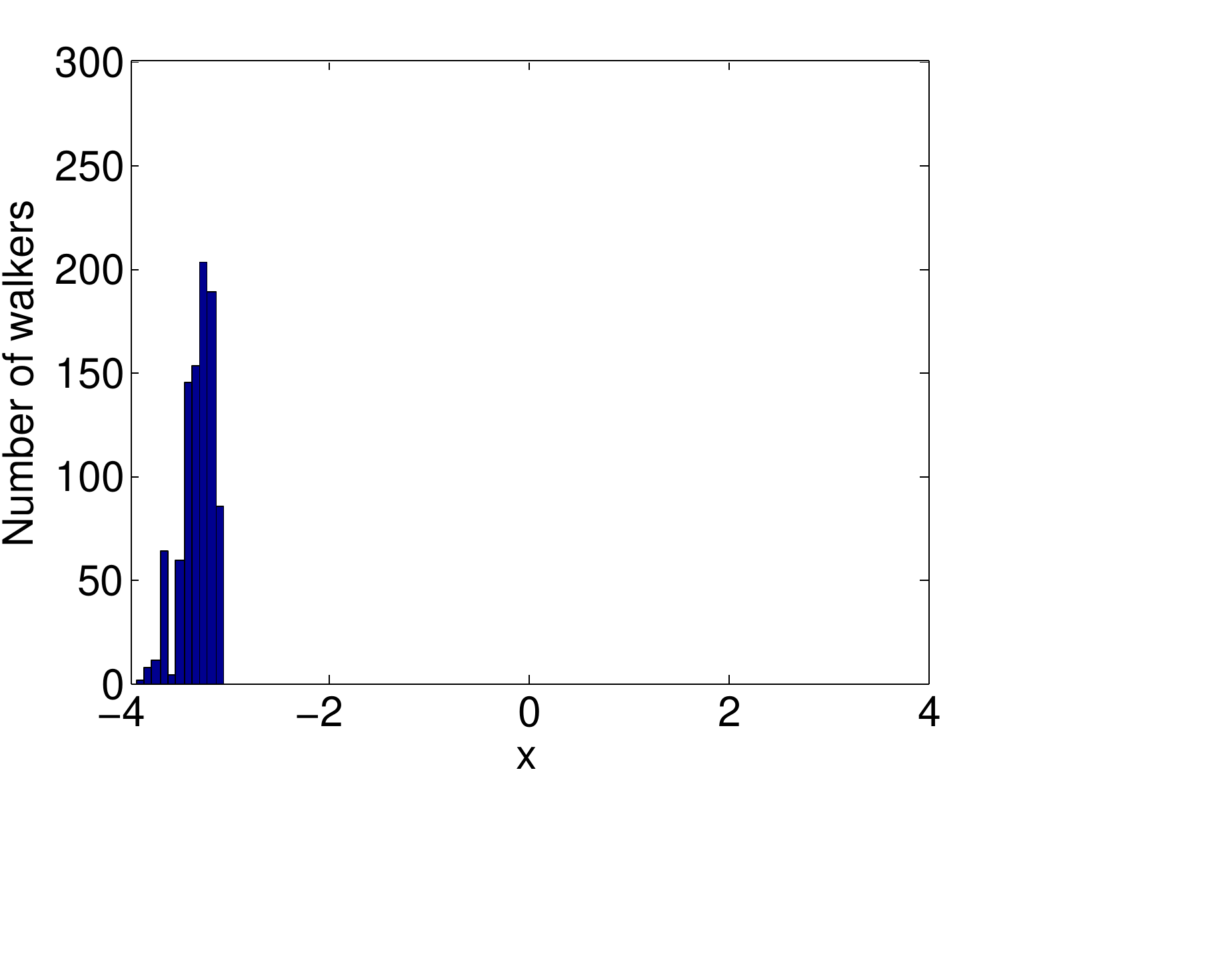}};

\node at (-11,1.8) {a)};
\node at (-4,1.8) {b)};
\node at (-11,-3.2) {c)};
\node at (-4,-3.2) {d)};

\end{tikzpicture}
\vskip -1.5cm
\caption{Distribution of walkers after the first time step, $T = \pi/4$, from the positive real ground state \( \psi_+(a) \) of ODHO, followed by cancellation. All four components of the wave function, a) positive real ($N \approx 6.27 \times 10^7$), b) negative imaginary ($N \approx 6.26 \times 10^7$),  c) negative real ($N \approx 2.0 \times 10^3$) and d) positive imaginary ($N \approx 0.9 \times 10^3$) walkers.  Note the different scaling on the vertical axes of the latter two.  Notations are the same as in Fig.~\ref{XP&XN}.}
\label{osc1}
\end{figure}

For the stationary ground state dynamics ($�E = 1�$), in each time step we expect to see the rotation of the phase factor $\exp(-i E t / \hbar) = \exp(-i  t )$, only, without any change in the absolute value of the wave function. 
Thus, the dynamics is expected to be simple oscillation of the real and imaginary parts of the ground state wave function in a phase difference of $\pi / 2$. 
The initial phase is chosen to be zero at $T_{0} = 0$, {\it i.e.}, \( \psi(0) = \psi_+(a) \) as before.  We start with $ N(a) = 10^7$ and run the simulation with the exact kernel (\ref{HOkernel}), time steps $t = \pi/4$ and duplicating walkers in $x_a$ enough so that after the cancellation $N(b) \ge N(0)$. Fig \ref{osc1} shows the distribution of remaining walkers after the first time step, $T = \pi/4$. 

As expected, we find the same copy of the starting gaussian as the positive real and imaginary parts and small remnants of incomplete cancellation in  both opposite sign parts, as a numerical error.  Here, with the walker touch parameter $\delta = 0.01$, the remaining opposite sign walkers are less than the proper walkers with a factor smaller than $10^{-4}$.  Thus, the cancellation is almost perfect.

\begin{figure}
\vskip-3cm
\centering
\begin{tikzpicture}

\node at (-9,0) {\includegraphics[trim=3cm 5cm 0.1cm 8cm, clip=true,scale=0.43]{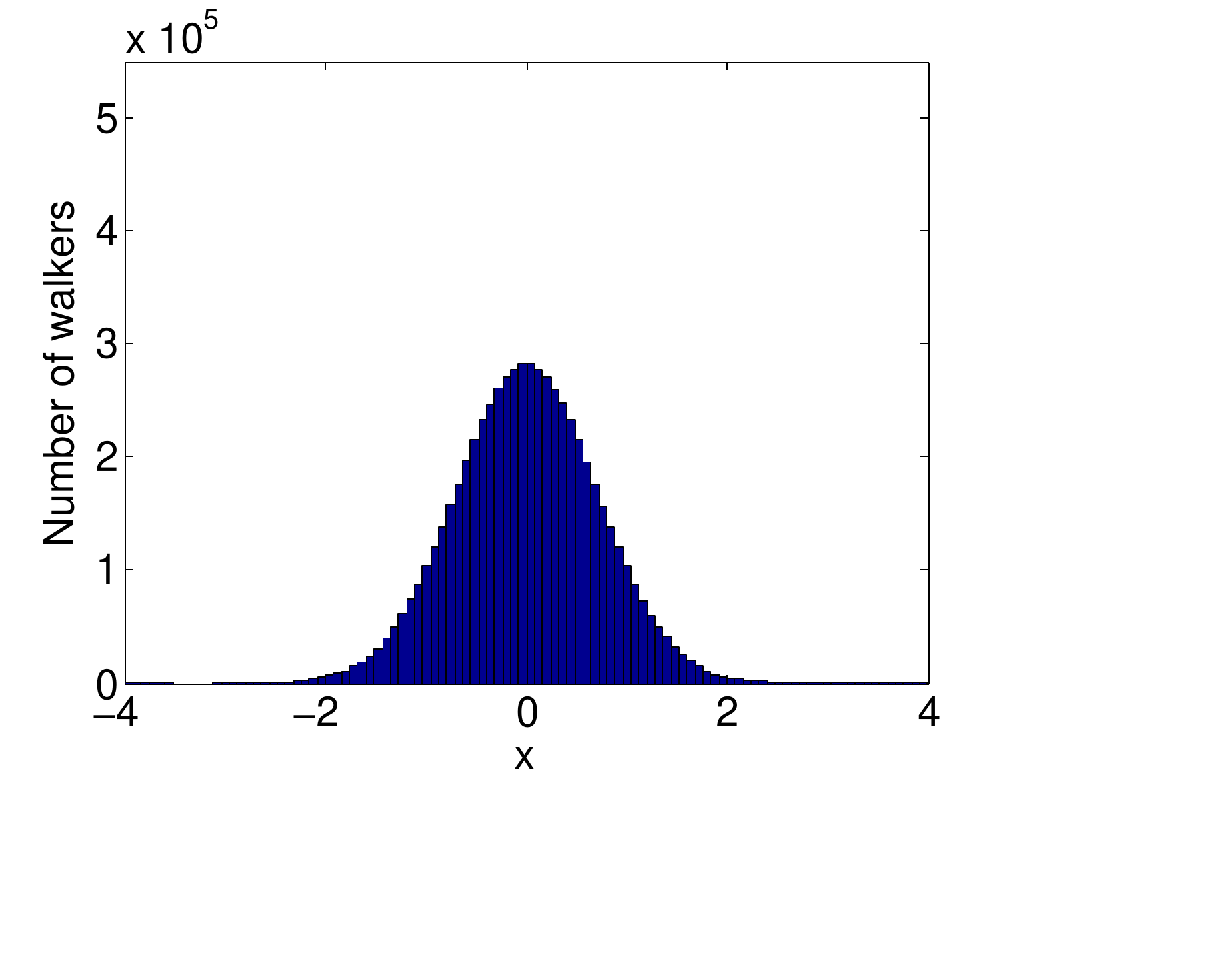}};
\node at (-2,0) {\includegraphics[trim=3cm 5cm 0.1cm 8cm, clip=true,scale=0.43]{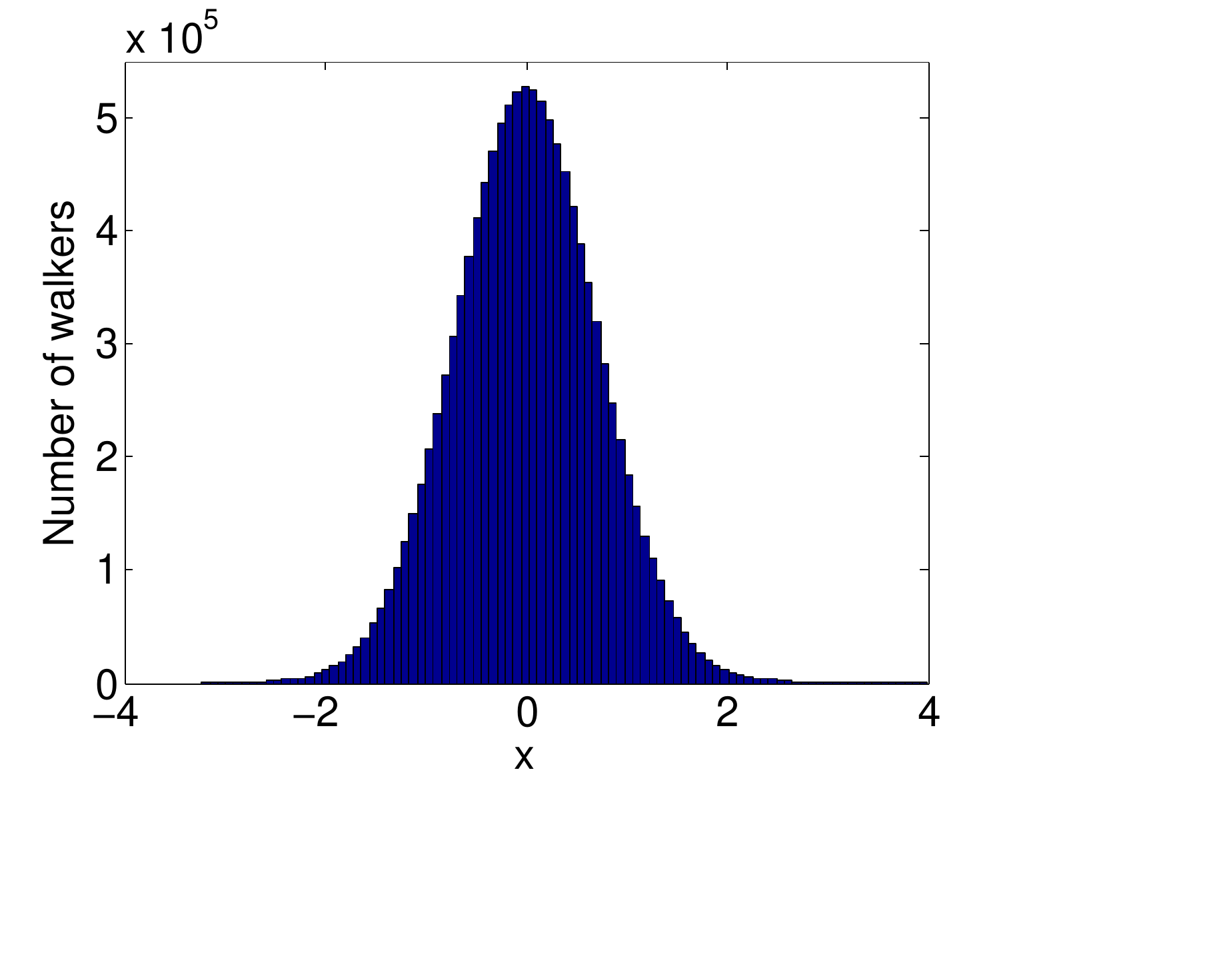}};
\node at (-9,-5) {\includegraphics[trim=3cm 5cm 0.1cm 8cm, clip=true,scale=0.43]{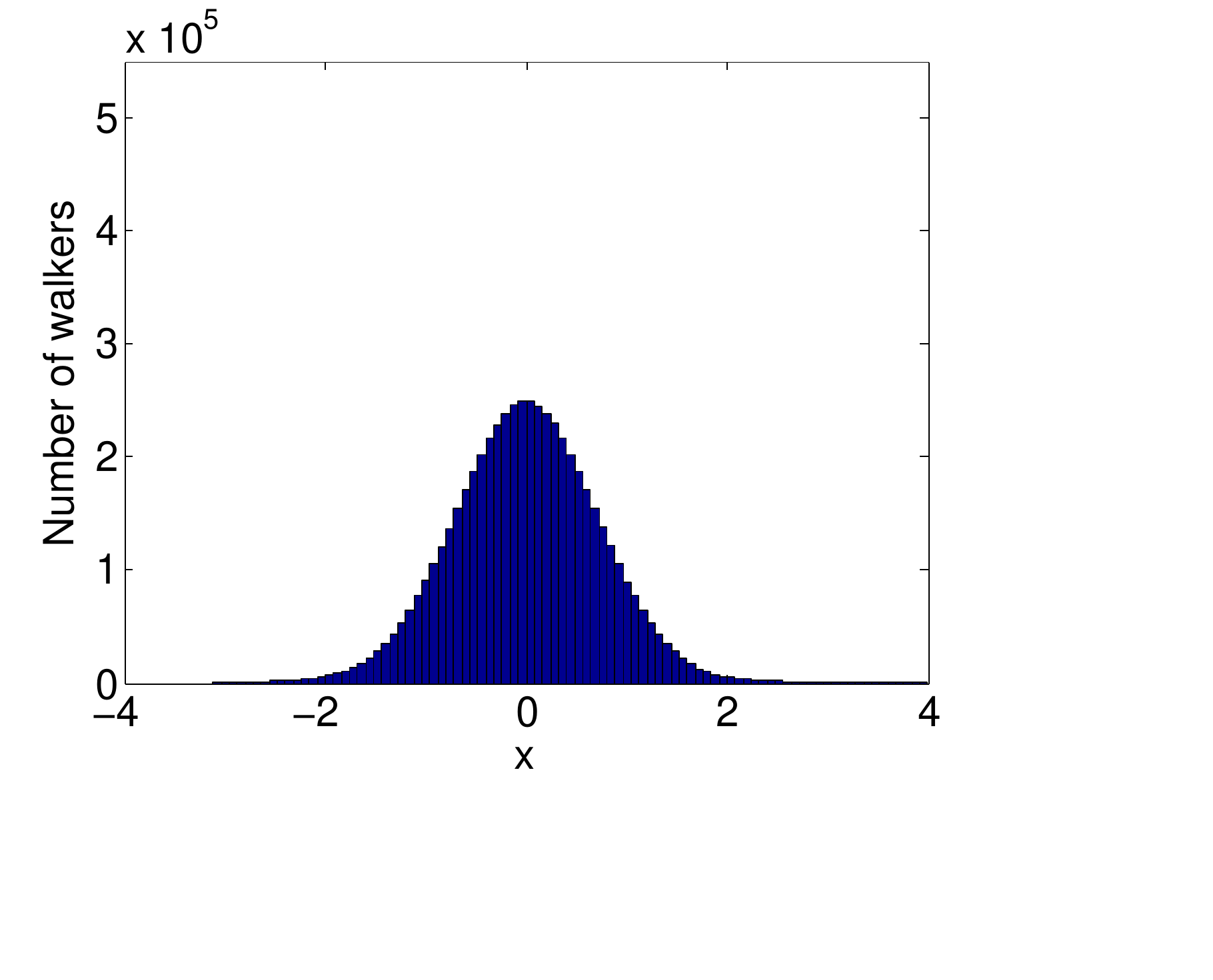}};
\node at (-2,-5) {\includegraphics[trim=3cm 5cm 0.1cm 8cm, clip=true,scale=0.43]{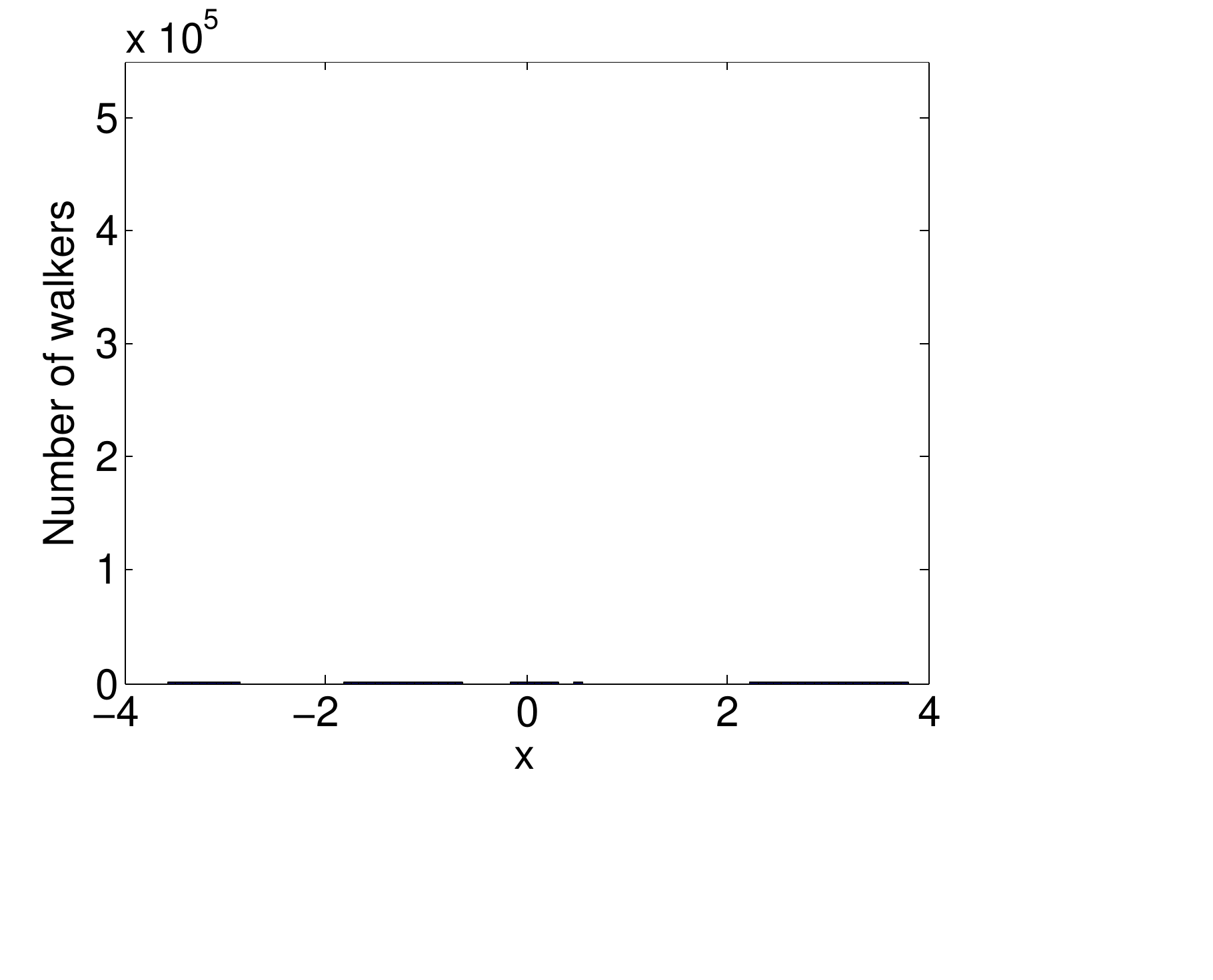}};

\node at (-11,1.8) {a)};
\node at (-4,1.8) {b)};
\node at (-11,-3.2) {c)};
\node at (-4,-3.2) {d)};

\end{tikzpicture}
\vskip -1.5cm
\caption{Distribution of negative imaginary walkers at a) $T = \pi/4$, b) $T = 2 \pi/4$, c)  $T = 3 \pi/4$ and d) $T = 4 \pi/4$ in the dynamics started in Fig.~(\ref{osc1}). Notations are the same as in Fig.~\ref{osc1}.}
\label{osc2}
\end{figure}
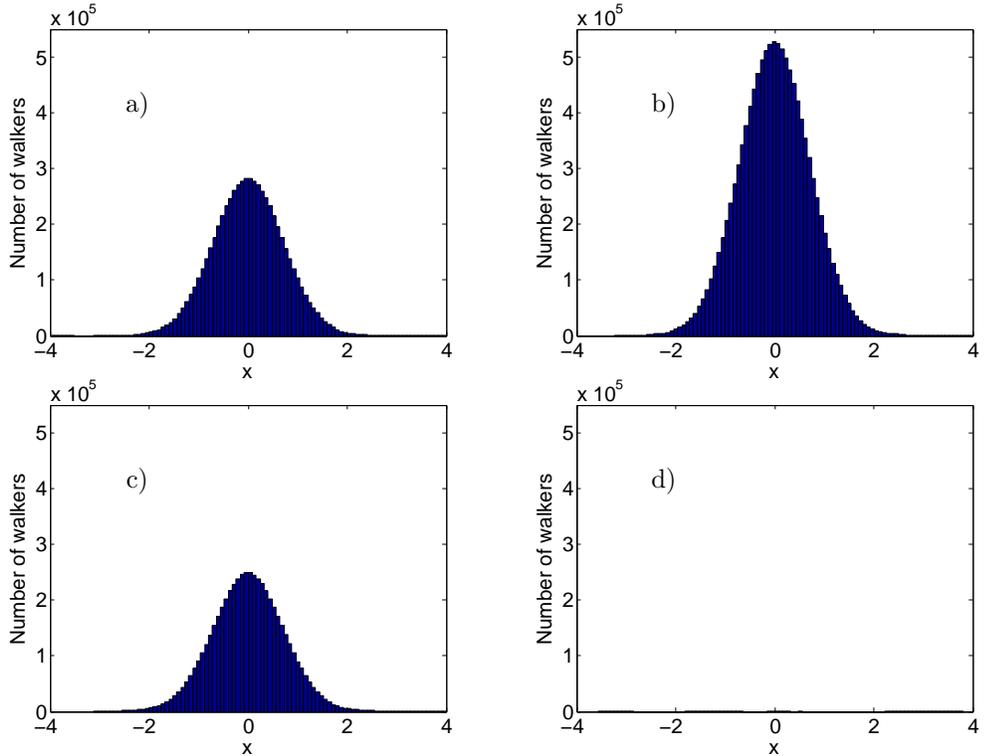

In Fig. \ref{osc2} we show the negative imaginary part of the wave function from further simulation, at times $ T = \pi/4,  2 \pi/4, 3 \pi/4, $ and $ 4 \pi/4$.  Clearly, the evolution is correct and at  $ T = \pi$ the wave function is purely real and negative with zero imaginary contribution.

%

\subsection{Evaluation of observables and eigenenergies}

Evaluation of transient expectation values of local operators, like multiplicative potential energy faces the same problem as with the $\tau$DMC, the wave function is given by the walker density, only.  Application of operators on the wave function or even finding the square of the wave function $�\psi^* \psi�$ numerically is not straightforward.  In our earlier studies we have demonstrated, that for $\tau$DMC one can easily evaluate the complex valued wave function of the system at each $\tau$DMC walker by using our direct real time path integral (RTPI) approach \cite{paper1}.  The RTPI time step is heavy to calculate, and therefore, could be restricted only to a few $\tau$DMC iteration steps, where needed.

Now, the RTPI can be used together with tDMC similarly as with $\tau$DMC in cases, where the wave function is purely real or imaginary.  This becomes relevant and useful with eigenstates and incoherent dynamics, in the next section.

With the eigenstates we should be able to monitor the phase factor of the wave function to find the corresponding eigenenergies.  Now, we cannot evaluate the local energy for each walker as can be done with RTPI \cite{paper1}.  However, we can evaluate the change in the ratio of the number of real and imaginary walkers to approximate the average collective change in the phase factor.  Thus, for the eigenenergy we write
\begin{equation} \label{Energyl}
E = -\frac{\theta \hbar}{t} = -\tan^{-1} \left( \frac{\psi_{\rm Im}}{\psi_{\rm Re}} \right) \frac{\hbar}{t} \approx \tan^{-1} \left( \frac{N(x_{\mp i})}{N(x_{\pm})} \right) \frac{\hbar}{t} . 
\end{equation}                                                                                             

For this to be accurate the time step should be short enough that the phase angle $\theta$ is small, but also, the ratio $N(x_{\mp i}) / N(x_{\pm})$ should be close to one so that the noise effect is minimised.  Furthermore, one should keep track of the quadrants of the complex plane and corresponding changes of sign, where relevant.

If the wave function is not an eigenstate but a superposition, for a short time step and small angle we can approximate
\begin{equation}
-\frac{\theta \hbar}{t} =   -\tan^ {-1} \left( \frac{\sum_i c_i \sin(\theta_i) }{\sum_i c_i \cos(\theta_i)} \right) \frac{\hbar}{t}\approx -\tan^ {-1}(\frac{\sum_i c_i \theta_i }{\sum_i c_i})\frac{\hbar}{t} \approx \frac{\sum_i c_i E_i }{\sum_i c_i} = E
\end{equation}                                                                                            
where the sum goes over the eigenstates with contributions  $c_i$.

%
%
%


%

\section{Incoherent propagation}

Earlier, we have developed the RTPI for coherent quantum dynamics and another RTPI version with incoherent dynamics for finding the eigenstates and energies of a system \cite{paper1}.  The incoherent dynamics is kind of quantum Zeno propagation, where the wave function is kept real.  In numerical simulation this can be accomplished by collapsing the complex wave function to a real one after each short time step.  In practise, the complex wave function is projected onto the real values by dropping off the imaginary part \cite{paper1}.

\subsection{Finding excited eigenstates}

The $\tau$DMC simulation converges to the lowest eigenstate (ground state) by adjusting the potential zero reference parameter $ E_T $ in Eq.~(\ref{DMCkernel}) to the lowest eigenvalue.  The convergence is usually unstable and needs continuous regulation with $ E_T $.  Recently, we have shown that the incoherent propagation of real time path integral dynamics RTPI drives the system to an eigenstate, which is closest to the zero reference of the potential energy \cite{paper1}.  Furthermore, the convergence is stable and does not need careful adjustment of potential zero reference.

Here too, we can insert the zero reference parameter $ E_T $ into the Eq.~(\ref{trottpropag}) and use it to choose the energy, for which we want to find the closest excited state.  Also, we can scan the parameter $ E_T $ to find all eigenstates within a given range.



%

%

\begin{figure}
\vskip -2cm
\includegraphics[trim=0.1cm 4cm 0.1cm 5cm, clip=true,scale=0.43]{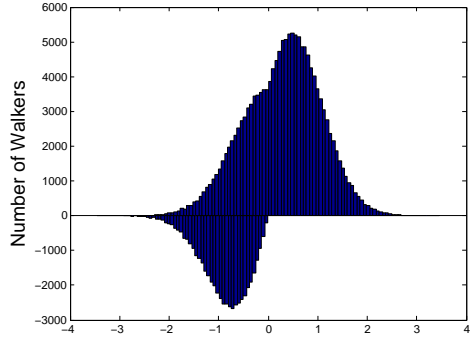}
\vskip -1.5cm
\caption{Positive ($N = 150 \times 10^3$) and negative walkers ($N = 50 \times 10^3$) of the superposition of 1st excited and the ground state ($N = 100 \times 10^3$ each).  Other notations are the same as in above figures.}
\label{Nex&Mix}
\end{figure}

Fig. \ref{Nex&Mix} shows a superposition of walkers of the real ground state and those of the real first excited state.  We see that the representation of the superposition is not unique, but calls for cancellation of positive and negative walkers.  However, we demonstrate robustness of the incoherent tDMC by starting with this initial wave function and run $ 100 $ time steps of length $t = 0.1$ with $10^6$ walkers.  The zero reference is set as $E_T = 0$.

\begin{figure}
\vskip -2cm
\includegraphics[trim=0.1cm 8cm 0.1cm 5cm, clip=true,scale=0.43]{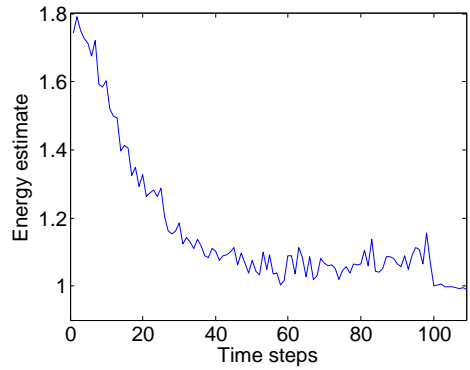}
\caption{Estimated energy that demonstrates convergence starting from the superposition of the 1st excited state and ground state in incoherent tDMC ending to the ground state.  The exact ground state eigenenergy is one, $E = 1$. $ N \approx 10^6$, and $t = 0.1$ for the first 100 time steps and then $t = 0.8$.}
\label{Energy}
\end{figure}

We monitor the eigenenergy from Eq.~(\ref{Energyl}) in Fig.~\ref{Energy}.  The exact value $E = 1$ is expected.  It can be seen that the convergence has been achieved in about 60 time steps to about $ E = 1.1 $.  Thus, there is some systematic error left, which we trace coming from the short time step.  With a too short time step false positive imaginary walkers appear, although all correct imaginary contribution should be negative.  This seems to relate also with the size of the domain, 8 atomic units.  Now, increasing the time step to $t = 0.8$ after 100 steps improves the energy estimate as clearly seen in the last ten time steps.  Then, the energy estimate from simulation is  $ 0.9974 \pm 0.0030$ (2 SEM).


%

%

\begin{figure}
\includegraphics[trim=0.1cm 5cm 0.1cm 5cm, clip=true,scale=0.43]{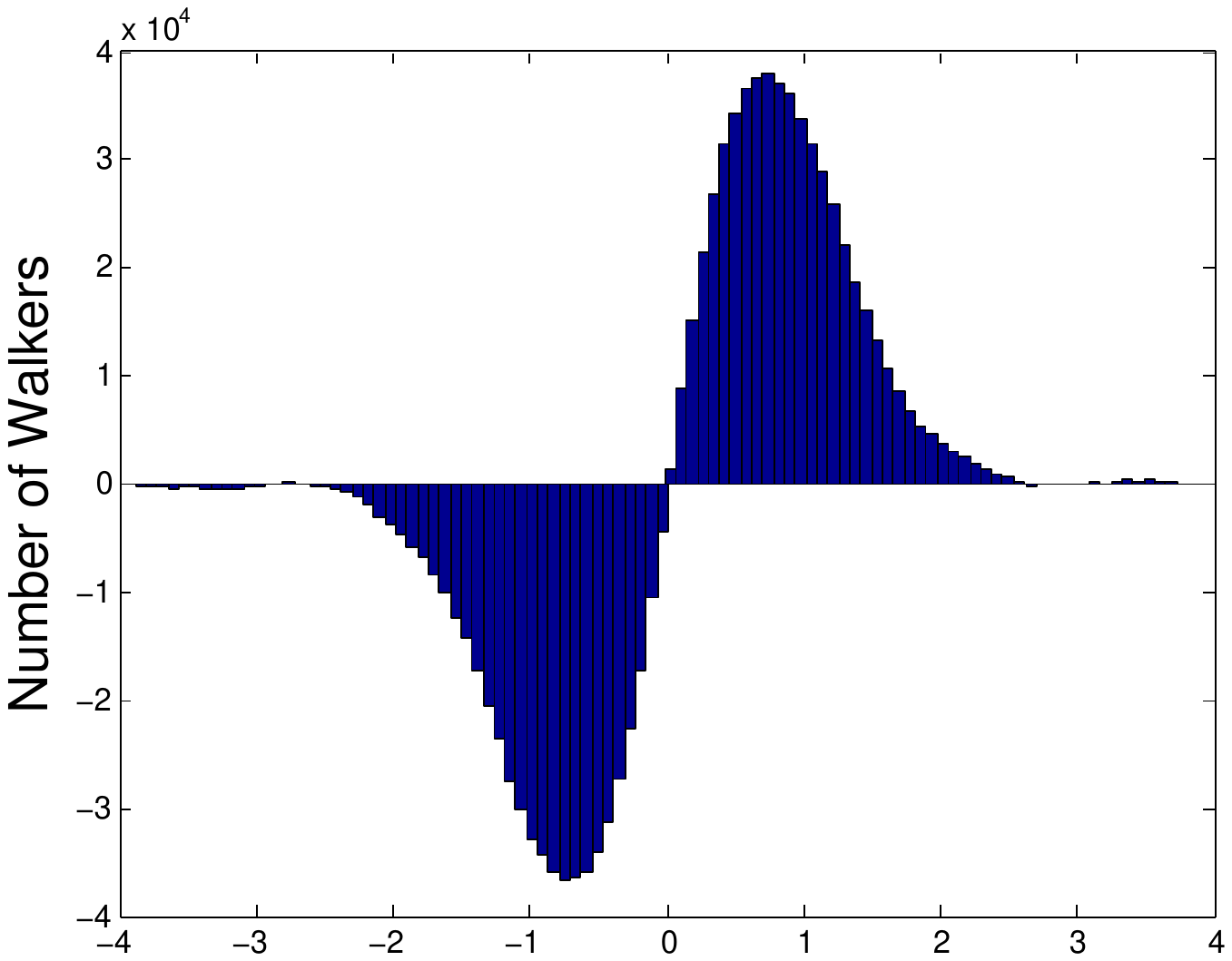}
\vskip -1cm
\caption{Distribution of positive ($N \approx 0.57 \times 10^6$) and negative ($N \approx 0.56 \times 10^6$) real walkers after the system has converged to its 1st excited state.}
\label{XNex}
\end{figure}

Finally, we search for the first excited state by using the incoherent propagation and starting from the same initial superposition state shown in Fig. \ref{Nex&Mix}.  Now, the potential zero reference is set as $E_T = 2.5$ and we expect to find the eigenenergy of 3.

By using a time step $t = \pi/12$ the first excited state is found as shown in Fig.~\ref{XNex} and the eigenenergy becomes as $3.0199 \pm 0.0076$ (2 SEM). Fig. \ref{XNex} shows the distribution of walkers after 100 timesteps to the convergence.  As the figure shows, the node of the wave function is clear and sharp.  By fitting to the histogram we get $0.0191$, which is close to the exact value of 0.

This approach may be one of the practical ways to locate nodal surfaces for other QMC methods like $\tau$DMC, and thus, give help in finding the practical solutions to the fermion sign problem.

\section{Conclusions}

We have demonstrated how the  real-time path integral kernel $K(x_b,t_b;x_a,t_a)$, Eq.~(\ref{kernel}), can be used to evaluate the time evolution of a  wave function with an entirely new way: driving delocalised "diffusion" of Monte Carlo walkers.  Therefore, we call our new approach as real-time DMC or tDMC.  There is a transparent analogy with the conventional imaginary time DMC or $\tau$DMC, where a local kernel $G(x_b,\tau_b;x_a,\tau_a)$, Eq.~(\ref{DMCkernel}), drives ordinary like diffusion of walkers in imaginary time.  However, it should be noted that tDMC is based on the real time path integral formalism, but $\tau$DMC is not!

It had been suspected that the real time counterpart of $\tau$DMC can not be realised, because the oscillating complex valued $K$ delocalised in space is not capable of driving real time diffusion similarly as the everywhere positive and normalizable $G$ drives imaginary time diffusion.  It was known, of course, that the real time kernel can be used to evaluate the time-dependent wave function by using the Eq.~(\ref{realtime}) directly, which couples all walkers within a time step making the numerical calculations heavy.  For that and some other practical reasons we were the first to realise the Real Time Path Integral (RTPI) approach for such light particles as electrons \cite{paper1,paper2}.

Thus, our tDMC is a truly novel QMC method.  It incorporates the essential features of $\tau$DMC, and similarly, can be used to find the system ground state energy and wave function with accuracy depending on the computational capacity.  In addition, with tDMC one can find also the excited states and the wave function nodes.  The latter may turn out to be useful in practical solutions of the fermion sign problem.

The tDMC can be run for incoherent dynamics or coherent dynamics, the same way as RTPI.  The former is used to find the eigenstates, while the latter, for evaluation of the time evolution of a wave function.  Comparison of tDMC and RTPI in running quantum dynamics is interesting.  In RTPI the walker distribution is (or follows) the wave function, {\it i.e.}, it is essentially localised in the wave function.  This may restrict the wave function response to fast transient effects or tunneling to a region, where walkers do not exist.  The tDMC with the fully delocalised diffusion, instead, fills the whole considered space with excess walkers in each time step before cancelling of walkers takes place.  Thus, the propagation is fully delocalised in the whole space in the spirit of path integrals, though the actual wave function may remain relatively localised.  Thus, the time evolution immediately responds to any distant changes in the external potential and allows start of tunneling to a region, where the wave function is essentially zero.


As we consider this first study as a "proof of concept" for tDMC, we chose a transparent and well-known one-dimensional harmonic oscillator as the test bench for the demonstration.  Now, the tDMC remains to be tested with many-particle systems, where we do not expect any surprises but the same course that we had with the RTPI \cite{paper1,paper2}, recently.

\section*{Acknowledgements}

For computational resources we thank Techila Technologies and TCSC for their facilities at Tampere University of Technology, and also, services of the Finnish IT Center for Science (CSC).

%

\end{document}